\documentclass{paper}

\usepackage{amsmath}
\usepackage{graphicx}

\begin{document}

\title{Exact paraxial diffraction theory for polygonal apertures under Gaussian illumination}

\author{Riccardo Borghi\\
Dipartimento di Ingegneria, Universit\`a degli Studi ``Roma Tre''\\
email: riccardo.borghi@uniroma3.it
}

\maketitle



\begin{abstract}
Paraxial diffraction of monochromatic Gaussian beams by arbitrarily shaped polygonal apertures is analytically explored within the boundary diffraction wave theory
framework. Exact closed-form expressions of the diffracted wavefield are obtained, as well as an interesting connection between classical optics and probability theory.
\end{abstract}






\section{Introduction}
\label{Sec:Introduction}

Fresnel's diffraction is a cornerstone of classical optics since more than two century.
Gaussian beams are the basic model to describe the light emitted by real laser sources.
The present paper is aimed at providing {exact} analytical expressions for the optical wavefield produced, within Fresnel's approximation,  
when a Gaussian beam is diffracted by an arbitrarily shaped sharp-edge  polygonal aperture. 
Although analytical descriptions for plane-wave Fraunhofer diffraction
by polygonal apertures are available since many years~\cite{Smith/Marsh/1974,Ganci/1984,Komrska/1982},
the same cannot be said when the diffracted field has to be estimated in the near field, apart from a few notable exceptions~\cite{Forbes/Asatryan/1998,Huang/Christian/McDonald/2006,Naraga/Hermosa/2018}. 

We believe what is contained in the present paper could represent a useful mathematical tool  for further improving
our theoretical understanding of diffraction. 
At the same time, the availability of a complete and exact knowledge of diffraction under an incomparably more realistic model of illumination
than a plane wave should be viewed, from a practical point of view, as an important achievement.
The whole theoretical analysis described below is within the context of a  ``genuinely paraxial'' formulation of the  
Young-Maggi-Rubinowicz boundary diffraction wave (BDW) theory~\cite{Born/Wolf/1999}
and its generalization by  Miyamoto and Wolf~\cite{Miyamoto/Wolf/1962},
recently developed~\cite{Borghi/2015,Borghi/2016,Borghi/2017,Borghi/2018,Borghi/2019}.
In what follows  only the main ideas and results will be highlighted, with several mathematical steps and details having been  dropped.
We encourage the interested readers to go through the above quoted papers to get a more complete idea of the general context. 

\section{Theoretical analysis}
\label{Sec:TheoreticalAnalysis}
 
\begin{figure}[!ht]
\centerline{\includegraphics[width=8cm,angle=-90]{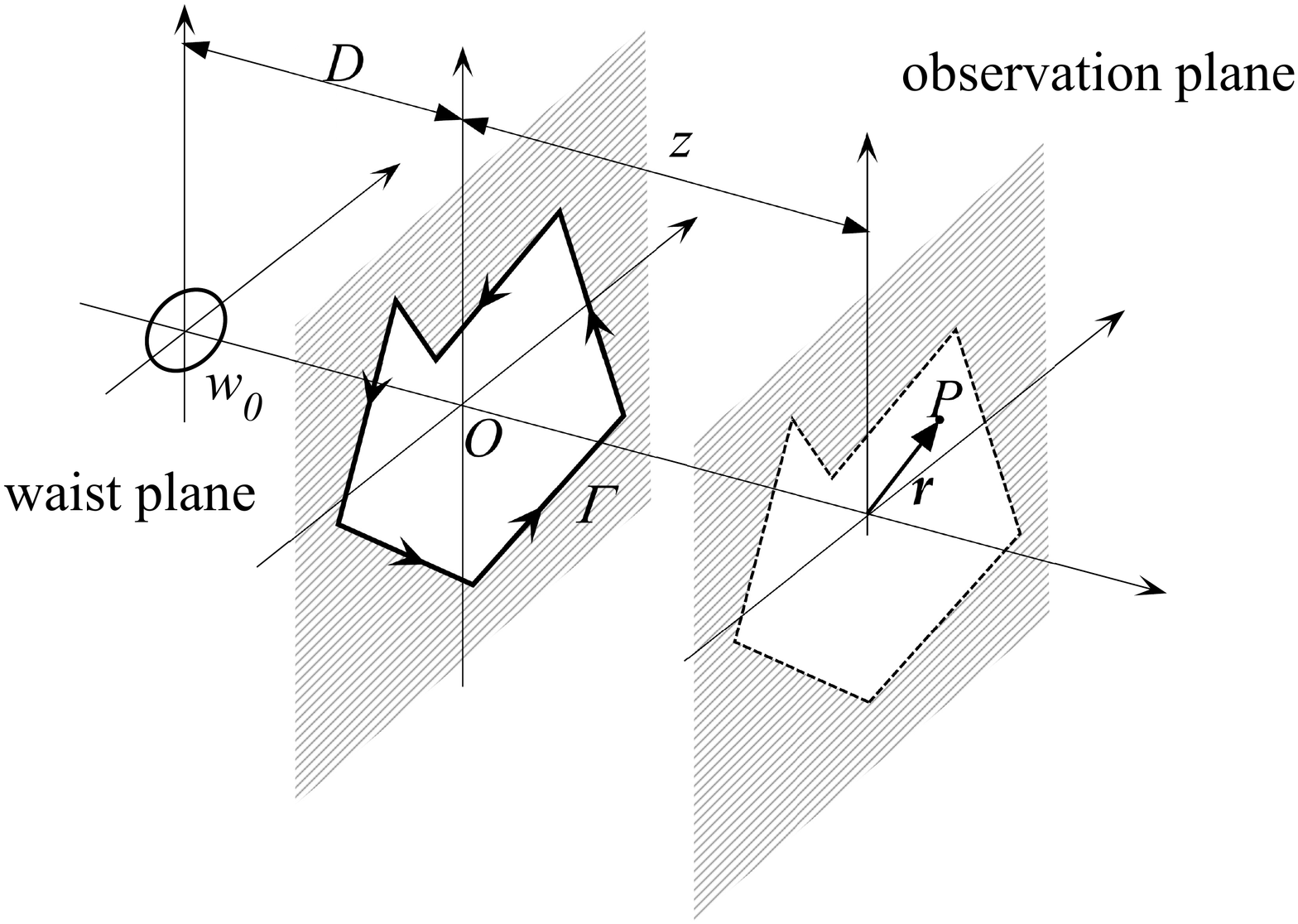}}
\caption{Geometry of  the problem.}
\label{Fig:Fresnel.1}
\end{figure}

The geometry of the problem is skecthed in Fig.~\ref{Fig:Fresnel.1}: 
a monochromatic Gaussian beam (wavenumber $k$, waist size $w_0$) orthogonally impinges  on an opaque 
transverse plane placed at a distance $D$ from its waist plane having a sharp-edge aperture $\mathcal{A}$
arbitrarily shaped as a $N$-side polygon. 
The problem to find the disturbance of the diffracted wavefield at the observation point $P \equiv (\boldsymbol{r};z)$
on a transverse plane at a distance $z>0$ from the aperture plane has been addressed from a rather general point of view in~\cite{Borghi/2019},
whose  main aspects will now be briefly resumed.
A key role is played by the following decomposition formula: 
\begin{equation}
\label{Eq:FresnelPropagatorConvolution.3.New.0.1}
\begin{array}{l}
\displaystyle
-\frac{\mathrm{i}\,k}{2\pi\,z}
\int_{\boldsymbol{\rho}\in\mathcal{A}}\,
\mathrm{d}^2\rho\,
\exp\left(\frac{\mathrm{i}k}{2z}\,|\boldsymbol{r}-\boldsymbol{\rho}|^2\right)\,=\,  \psi_G\,+\, \psi_{\rm BDW}\,,
\end{array}
\end{equation}
where the complex quantities $\psi_G$ and $\psi_{\rm BDW}$ are both expressed via suitable one-dimensional contour integrals defined
onto the aperture boundary $\Gamma=\partial\mathcal{A}$. On denoting the position of a typical point on  $\Gamma$
by $Q$, let  $Q=Q(t)$ be  a suitable parametrization of $\Gamma$, with $t$ being a real parameter ranging within a given interval. 

Then, on  introducing  the transverse vector  $\boldsymbol{R}=\overrightarrow{PQ}$, quantities $\psi_G$ and $\psi_{\rm BDW}$
can be evaluated through~\cite{Borghi/2019} 
\begin{equation}
\label{Eq:FresnelPropagatorConvolution.3.New.4}
\begin{array}{l}
\displaystyle
\psi_{G}\,=\,\frac{1}{2\pi}\,\oint_\Gamma\,\mathrm{d} t\, 
\dfrac{\boldsymbol{R}\times\dot{\boldsymbol{R}}}
{\boldsymbol{R}\cdot\boldsymbol{R}}\,,
\end{array}
\end{equation}
and
\begin{equation}
\label{Eq:FresnelPropagatorConvolution.3.New.5}
\begin{array}{l}
\displaystyle
\psi_{\rm BDW}\,=\,-\frac{1}{2\pi}\oint_\Gamma\,\mathrm{d} t\, 
\dfrac{\boldsymbol{R}\times\dot{\boldsymbol{R}}}
{\boldsymbol{R}\cdot\boldsymbol{R}}\,
\exp\left(\frac{\mathrm{i}k}{2z}\,\boldsymbol{R}\cdot\boldsymbol{R}\right)\,,
\end{array}
\end{equation}
respectively. 
Here $\dot{\boldsymbol{R}}$ denotes the derivative of $\boldsymbol{R}$ with respect to the parameter $t$, 
while the cross product must be intended as the sole $z$-component, being both vectors $\boldsymbol{R}$ and $\dot{\boldsymbol{R}}$ 
purely transverse. 
In~\cite{Borghi/2019} it was conjectured the above integral representations to be valid, in principle, also for {\em complex} values of the Cartesian 
coordinates of the observation point $P$. Accordingly,   the following recipe to retrieve the diffracted wavefield produced by the Gaussian beam in Fig.~\ref{Fig:Fresnel.1} was 
then proposed~\cite{Borghi/2019}: 
(i) multiply all Cartesian coordinates of $P$, i.e., the position vector $(\boldsymbol{r};z)$, by the following dimensionless complex factor
\begin{equation}
\label{Eq:FresnelPropagatorConvolution.3.New.5.1}
\begin{array}{l}
\displaystyle
\dfrac{1+\mathrm{i} D/L}{1+\mathrm{i} (z+D)/L}\,,
\end{array}
\end{equation}
with $L=kw^2_0/2$ being the incident Gaussian beam Rayleigh length;
(ii) evaluate both $\psi_G$ and $\psi_{\rm BDW}$ for the above complex values of coordinates through Eqs.~(\ref{Eq:FresnelPropagatorConvolution.3.New.4}) and~(\ref{Eq:FresnelPropagatorConvolution.3.New.5}); 
(iii)  multiply $\psi_G+\psi_{\rm BDW}$ by the wavefield obtained by letting the Gaussian beam to freely 
propagate from the waist plane up to the \emph{real} observation point $P\equiv (\boldsymbol{r};z)$ to  retrieve the diffracted wavefield. That is all.

\begin{figure}[!ht]
\centerline{\includegraphics[width=7cm,angle=-90]{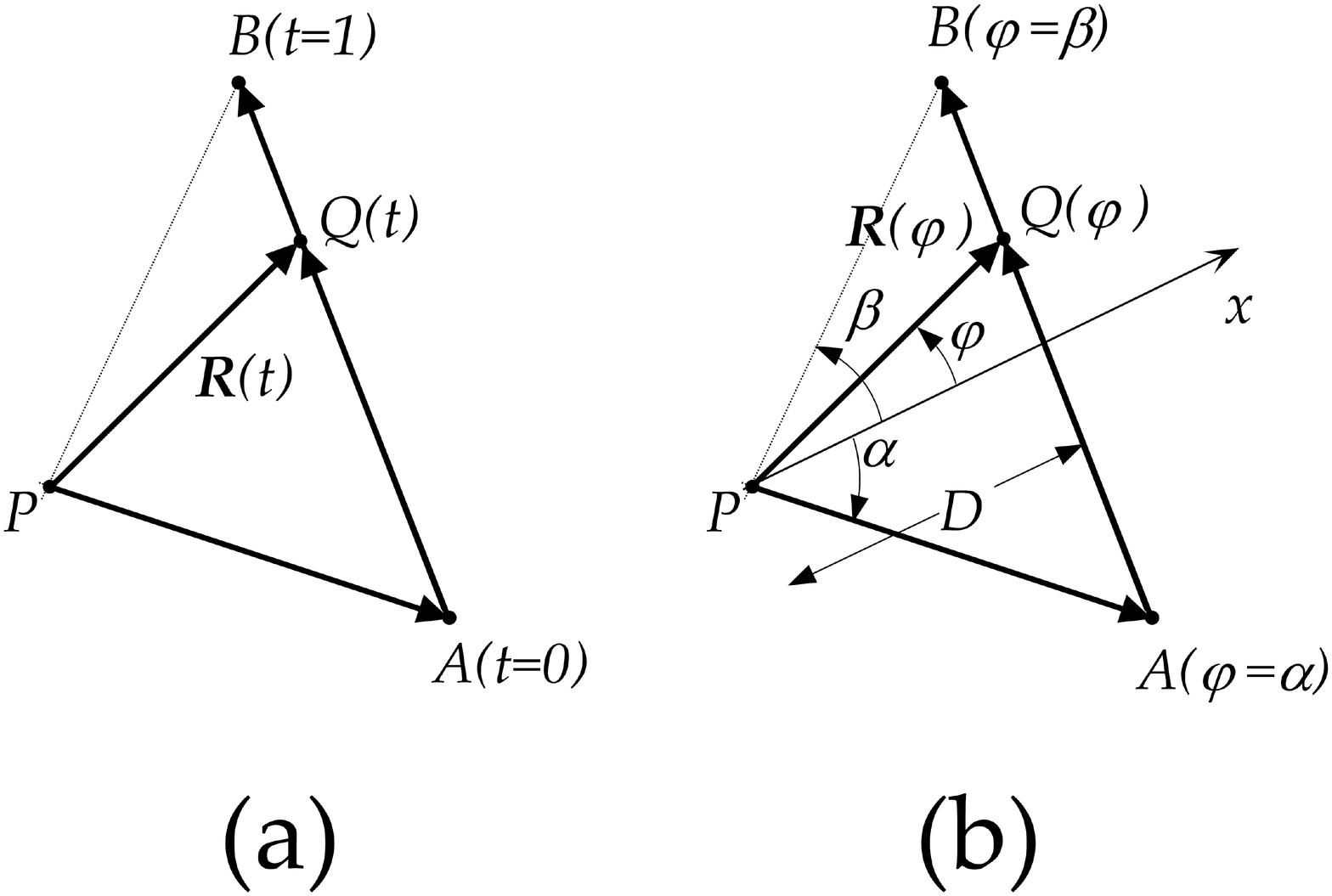}}
\caption{Parametrizations of the segment $AB$ for the evaluation of the contributions to 
the geometrical component $\psi_G(AB)$ (a) and of the BDW component $\psi_{\rm BDW}(AB)$ (b)
at the observation point $P$.}
\label{Fig:BDW.5}
\end{figure}

Now we are going to show how, when the aperture boundary $\Gamma$ is an arbitrarily shaped polygon, Eqs.~(\ref{Eq:FresnelPropagatorConvolution.3.New.4}) and~(\ref{Eq:FresnelPropagatorConvolution.3.New.5})
can be exactly expressed through analytical closed forms. 
The idea is simple: both $\psi_G$ and $\psi_{\rm BDW}$ can be written as the sum of a \emph{finite} number of contributions, each of them coming from one polygon side.
Accordingly, In this way the main problem reduces to evaluate Eqs.~(\ref{Eq:FresnelPropagatorConvolution.3.New.4}) and~(\ref{Eq:FresnelPropagatorConvolution.3.New.5}) for a typical segment, say $AB$, placed at the transverse 
plane at $z$, as sketched in Fig.~\ref{Fig:BDW.5}.

We start  with the contribution to the geometrical component, say $\psi_G(AB)$. To this aim, a natural parametrization of the segment is
(see Fig.~\ref{Fig:BDW.5}a)
\begin{equation}
\label{Eq:FresnelPropagatorConvolution.3.New.5.1.1}
\begin{array}{l}
\displaystyle
\boldsymbol{R}(t)\,=\,\overrightarrow{PA}\,+\,\overrightarrow{AQ}\,=\,\overrightarrow{PA}\,+\,\overrightarrow{AB}\,t\,,\qquad\qquad t \in [0,1]\,,
\end{array}
\end{equation}
so that $\dot{\boldsymbol{R}}\,=\,\overrightarrow{AB}$. On substituting from Eq.~(\ref{Eq:FresnelPropagatorConvolution.3.New.5.1.1})
into Eq.~(\ref{Eq:FresnelPropagatorConvolution.3.New.4}) we obtain at once
\begin{equation}
\label{Eq:FresnelPropagatorConvolution.3.New.5.1.2}
\begin{array}{l}
\displaystyle
\psi_G(AB)\,=\,\dfrac 1{2\pi}\,\dfrac{\overrightarrow{PA}\times\overrightarrow{AB}}{\overline{AB}^2}\,
\int^1_0\,\dfrac{\mathrm{d} t}{t^2\,+2\,\eta\,t\,+\,\chi^2}\,,
\end{array}
\end{equation}
where
\begin{equation}
\label{Eq:FresnelPropagatorConvolution.3.New.5.1.3}
\begin{array}{lcr}
\displaystyle
\eta\,=\,\dfrac{\overrightarrow{PA}\cdot \overrightarrow{AB}}{\overline{AB}^2}\,,
&
\chi^2\,=\,\dfrac{\overrightarrow{PA}\cdot \overrightarrow{PA}}{\overline{AB}^2}\,.
\end{array}
\end{equation}
Despite its apparent simplicity, the evaluation of the integral in Eq.~(\ref{Eq:FresnelPropagatorConvolution.3.New.5.1.2})
requires some care since it has to be computed, for what it was said above, when $P$ attains complex values of its coordinates,
say $(\boldsymbol{r}\,\exp(\mathrm{i}\phi);z\,\exp(\mathrm{i}\phi))$.  
Without detailing all mathematical steps, the following closed-form  formula can be established:
\begin{equation}
\label{Eq:FresnelPropagatorConvolution.3.New.5.1.4}
\begin{array}{l}
\displaystyle
\int^1_0\,\dfrac{\mathrm{d} t}{t^2\,+2\,\eta\,t\,+\,\chi^2}  
\displaystyle
\,=\,
\dfrac 1{2\sqrt{\eta^2\,-\,\chi^2}}\,\sum_{\sigma=\pm 1}\,\sigma\,\log\left(1+\dfrac{\eta+\sigma\sqrt{\eta^2-\chi^2}}{\chi^2}\right)\,,
\end{array}
\end{equation}
which, together with Eqs.~(\ref{Eq:FresnelPropagatorConvolution.3.New.5.1.2}) and~(\ref{Eq:FresnelPropagatorConvolution.3.New.5.1.3}),
gives the exact representation of the geometrical wavefield contribution due to the segment $AB$. As an easy check, it is an academic 
exercise to prove that, for \emph{real} values of $\eta$ and $\chi$, corresponding to plane-wave illumination, Eqs.~(\ref{Eq:FresnelPropagatorConvolution.3.New.5.1.2})-(\ref{Eq:FresnelPropagatorConvolution.3.New.5.1.4}) 
give at once
\begin{equation}
\label{Eq:FresnelPropagatorConvolution.3.New.5.1.5}
\begin{array}{l}
\displaystyle
\psi_G(AB)\,=\,\dfrac {\beta\,-\,\alpha}{2\pi}\,,
\end{array}
\end{equation}
where the sign of angles $\alpha$ and $\beta$ is positive for counterclockwise rotations (see Fig.~\ref{Fig:BDW.5}b for the meaning of $\alpha$ and $\beta$).

The evaluation of the contribution to the BDW component of the diffracted wavefield coming from the  segment $AB$, say $\psi_{\rm BDW}(AB)$,
 is more cumbersome, since involves the use of a special function which is well known in statistics but still unknown in optics. 
To evaluate this contribution, the integration parameter into Eq.~(\ref{Eq:FresnelPropagatorConvolution.3.New.5}) has to be changed 
from $t$ (see Fig.~\ref{Fig:BDW.5}a) to the polar angle $\varphi$ shown in Fig.~\ref{Fig:BDW.5}b. Let $D$ denote the distance between $P$ and $AB$. 
On introducing the Cartesian axis $x$ whose origin coincides with $P$ and which is orthogonal to the $AB$ direction,  the 
polar representation of the vector $\boldsymbol{R}$ is given by $R(\varphi)=D/\cos\varphi$, with $\varphi\in[\alpha,\beta]$.
Accordingly, the contribution of $AB$ to the BDW wavefield turns out to be
\begin{equation}
\label{Sec:BDW.3}
\begin{array}{l}
\displaystyle
\psi_{\rm BDW}(AB)\,=\, -\frac {1}{2\pi}\,
\int_\alpha^\beta\,
\exp\left(\frac {\mathrm{i} u/2}{\cos^2\varphi}\right)\,\mathrm{d}\varphi\,,
\end{array}
\end{equation}
where the dimensionless parameter $u=kD^2/z$ has been introduced.
Note that the distance $D$ in Fig.~\ref{Fig:BDW.5}b can formally be analytically expressed as 
\begin{equation}
\label{Sec:BDW.9.0}
\begin{array}{l}
\displaystyle
D\,=\,\frac{\overrightarrow{PA}\,\times\,\overrightarrow{AB}}{\overline{AB}}\,,
\end{array}
\end{equation}
where use has been made of $\overrightarrow{PB}\,=\,\overrightarrow{PA}\,+\,\overrightarrow{AB}$.
To evaluate the integral in Eq.~(\ref{Sec:BDW.3}), it  is then sufficient to make the variable change $\tau=\tan\varphi$, which yields
\begin{equation}
\label{Sec:BDW.3.1.1}
\begin{array}{lc}
\displaystyle
\psi_{\rm BDW}(AB)\,=\, 
T\left(\sqrt{-\mathrm{i}\,u},\tan\alpha\right)\,-\,T\left(\sqrt{-\mathrm{i}\,u},\tan\beta\right)\,,
\end{array}
\end{equation}
where the symbol $T(a,b)$ denotes the so-called \emph{Owen T-function}~\cite{Owen/1956},  defined as
follows:
\begin{equation}
\label{Sec:BDW.3.9}
\begin{array}{l}
\displaystyle
T(a,b)\,=\,\frac 1{2\pi}\,
\int_0^{b}\,\frac{\exp[-\frac {a^2}2(1+\tau)^2]}{1+\tau^2}\,\mathrm{d}\tau\,.
\end{array}
\end{equation}
For {real} values of  both  parameters, the quantity $T(a,b)$ gives the probability that  $\{X>a$ and $0<Y<b X\}$,
with $X$ and $Y$ being two statistically independent  normal random variables~\cite{Owen/1956}. 
For our purposes it is important to note that Owen's function $T(a,b)$ can be analytically continued to  complex values of both arguments.
A precious work about the main properties of the Owen function (including integral and series representations, 
connections with other special functions, evaluation of integrals and series containing the $T$-function) has recently 
been published~\cite{Brychov/Savischenko/2016}.
Readers are encouraged to go through this nice work to familiarize with the Owen function. 
Analytical expressions of both $\tan\alpha$ and $\tan\beta$ in terms of the positions of points $A$, $B$, and $P$  can also be found through elementary geometry, which yields
\begin{equation}
\label{Sec:BDW.9}
\left\{
\begin{array}{l}
\displaystyle
\tan\alpha\,=\,\dfrac{\overrightarrow{PA}\,\cdot\,\overrightarrow{AB}}{\overrightarrow{PA}\,\times\,\overrightarrow{AB}}\,,\\
\\
\displaystyle
\tan\beta\,=\,\dfrac{\overrightarrow{PB}\,\cdot\,\overrightarrow{AB}}{\overrightarrow{PA}\,\times\,\overrightarrow{AB}}\,.
\end{array}
\right.
\end{equation}
Equations~(\ref{Eq:FresnelPropagatorConvolution.3.New.5.1.2}) - (\ref{Eq:FresnelPropagatorConvolution.3.New.5.1.4}),
together with Eqs.~(\ref{Sec:BDW.9.0}) - (\ref{Sec:BDW.9}),  are the main result of the present paper. Through them,
paraxial diffraction of Gaussian beams by polygonal sharp-edges apertures finds an exact 
solution, as promised at the beginning.

In the rest of the paper a single but significant example of application of the theory previously exposed will be illustrated.
Such example received a considerable attention, even recently~\cite{Huang/Christian/McDonald/2006,Stahl/Gbur/2016,Rivera/Galvin/Steinforth/Eden/2018,Sroor/Naidoo/Miller/Nelson/Courtial/Forbes/2019,Rocha/Amaral/Fonseca/Jesus-Silva/2019}. 
\begin{figure}[!ht]
\centerline{\includegraphics[width=6cm,angle=-90]{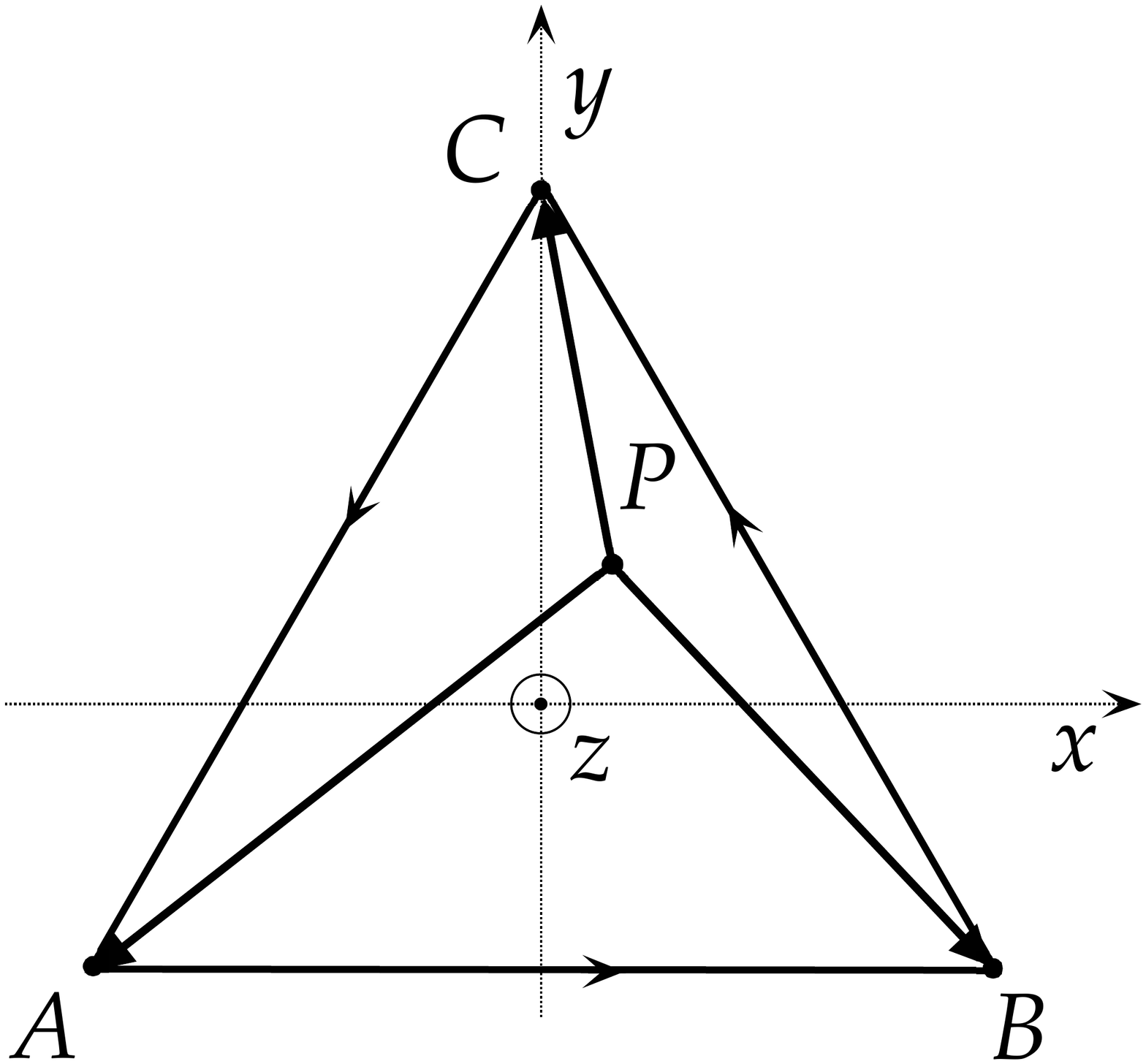}}
\caption{Geometry for Gaussian beam diffraction by equilateral triangular apertures.}
\label{Fig:TriangoloEquilateroGeometry}
\end{figure}
Figure~\ref{Fig:TriangoloEquilateroGeometry} depicts the geometry of the problem: a sharp-edge aperture shaped in the form of an equilater triangle is orthogonally illuminated by 
a Gaussian beam whose mean propagation axis passes through the triangle centre. It is worth introducing a Cartesian reference frame  $Oxyz$,
in which the coordinates of the triangle vertices are $A \equiv (-\sqrt 3,-1)$, $B \equiv (\sqrt 3,-1)$, and $C \equiv (0,2)$, where for simplicity the length unit has been chosen 
in such  way that each triangle side measures $2\sqrt 3$. Then, on denoting $(x,y)$ the Cartesian representation of the observation point $P$, 
Eqs.~(\ref{Eq:FresnelPropagatorConvolution.3.New.5.1.3}), (\ref{Sec:BDW.9.0}), and~(\ref{Sec:BDW.9}) allows all parameters $\eta$, $\chi^2$, $D$, $\tan\alpha$, and  $\tan\beta$ to 
be easily evaluated for all triplets $(A,B,P)$,  $(B,C,P)$, and $(C,A,P)$. For example, the parameters related to the triplet $(A,B,P)$ take on the following expressions:
\begin{equation}
\label{Sec:BDW.10}
\left\{
\begin{array}{l}
\displaystyle
\eta\,=\,-\dfrac{3+x\sqrt 3}{6}\,,\\
\\
\displaystyle
\chi^2\,=\,\dfrac{x^2+2x\sqrt 3+y(2+y)+4}{12}\,,\\
\\
\displaystyle
D\,=\,1+y\,,\\
\\
\displaystyle
\tan\alpha\,=\,-\dfrac{\sqrt 3+x}{1+y}\,,\\
\\
\displaystyle
\tan\beta\,=\,\dfrac{\sqrt 3-x}{1+y}\,.
\end{array}
\right.
\end{equation}
Similar expressions follow for the other two triplets $(B,C,P)$ and
$(C,A,P)$ but will not be given here.

The geometrical component  $\psi_G$ is first investigated. Several years ago Otis~\cite{Otis/1974} claimed a Gaussian beam impinging on a sharp-edge 
aperture would  produce a geometrical ``Gaussian shadow''  having the same shape as the aperture and scaled by the modulus of the factor in Eq.~(\ref{Eq:FresnelPropagatorConvolution.3.New.5.1}). Otis' conjecture implies the geometrical field defined by Eq.~(\ref{Eq:FresnelPropagatorConvolution.3.New.4}),
once evaluated at \emph{complex transverse positions} $\boldsymbol{r}\,\exp(\mathrm{i}\phi)$ must satisfy the following relation:
\begin{equation}
\label{Eq:FresnelPropagatorConvolution.3.New.3.1}
\begin{array}{l}
\displaystyle
\psi_G\,=\, 
\left\{
\begin{array}{lr}
1 & \boldsymbol{r} \in \mathcal{A}\,,\\
&\\
0 & \boldsymbol{r} \notin \mathcal{A}\,,
\end{array}
\right.
\end{array}
\end{equation}
\emph{irrespective} the value of $\phi$.
Equation~(\ref{Eq:FresnelPropagatorConvolution.3.New.3.1}) was rigorously proved for circular apertures  sharing the axial symmetry with the impinging Gaussian beam~\cite{Borghi/2019}.
In the same work  it was also argued  such a case to be the sole for which the Otis conjecture holds. The fascinating problem about such ``Gaussian shadows'' will surely be the subject of future investigations. In the present paper we shall limit to provide a single check about  the conjectures of~\cite{Borghi/2019}.
\begin{figure}[!ht]
\centerline{\includegraphics[width=8cm,angle=-90]{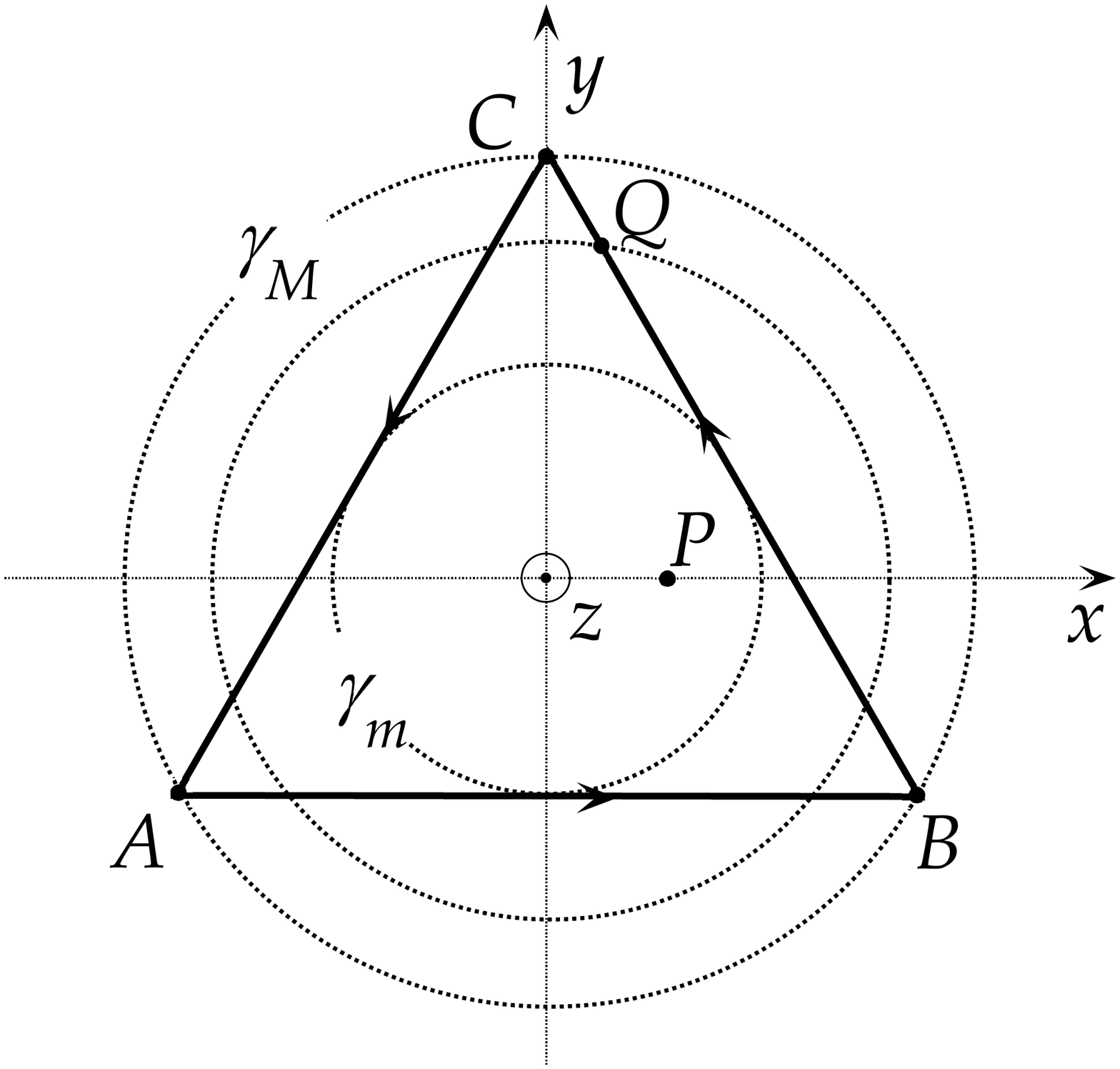}}
\caption{Geometry for checking the conjecture in~\cite{Borghi/2019}.}
\label{Fig:BorghiConjecture}
\end{figure}

\begin{figure}[!ht]
\centerline{\includegraphics[width=7cm,angle=-0]{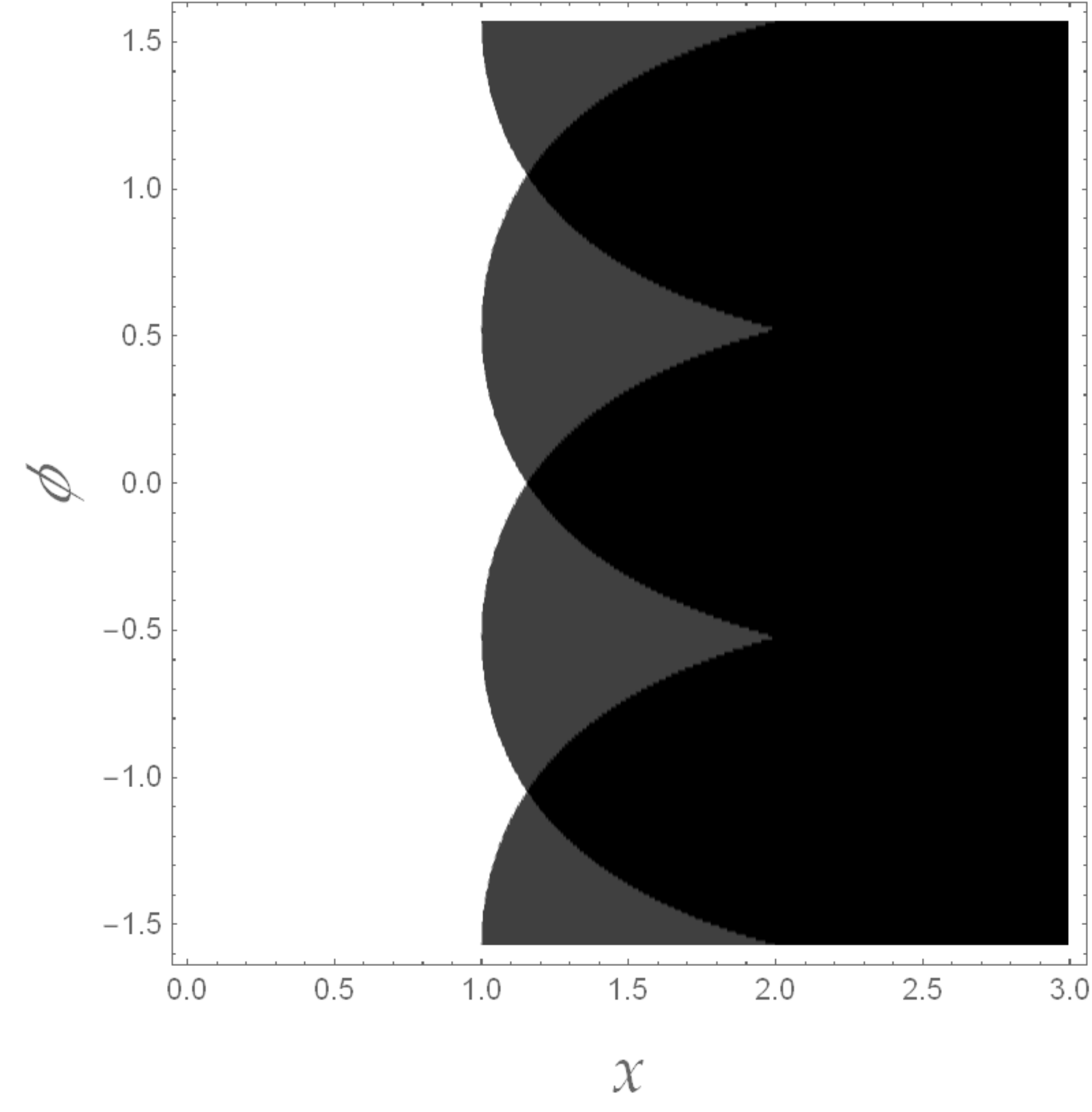}}
\caption{Two-dimensional map of the geometrical wavefield $\psi_G$ for the triangular geometry in Fig.~\ref{Fig:BorghiConjecture}, 
numerically evaluated at complex transverse observation points $(x\exp(\mathrm{i}\phi),0)$.}
\label{Fig:BorghiConjecture.2}
\end{figure}

Figure~\ref{Fig:BorghiConjecture} shows the geometry: the observation point $P$ belongs to the $x$-axis. For a given value
of $x$, the geometrical wavefield $\psi_G$ is then computed at $x\,\exp(\mathrm{i}\phi)$, with $\phi\in [-\pi,\pi]$. The result is shown in Fig.~\ref{Fig:BorghiConjecture.2}:
it turns out that $\psi_G=1$ within the  white region, $\psi_G=0$ within the  black region, and $\psi_G=1/2$ within the grey region. 

It is thus seen that Eq.~(\ref{Eq:FresnelPropagatorConvolution.3.New.3.1}) is not satisfied, except when the observation points is inside 
the inscribed circle $\gamma_m$, for which $\psi_G\equiv 1$, or when it is outside the circumscribed circle $\gamma_M$, for which $\psi_G\equiv 0$. 
Several others numerical trials, made at different observation points but not shown here, gave results in perfect agreement with what has been conjectured in~\cite{Borghi/2019}.
\begin{figure}[!ht]
\centering
\begin{minipage}[t]{10cm}
\includegraphics[width=4.8cm,angle=-0]{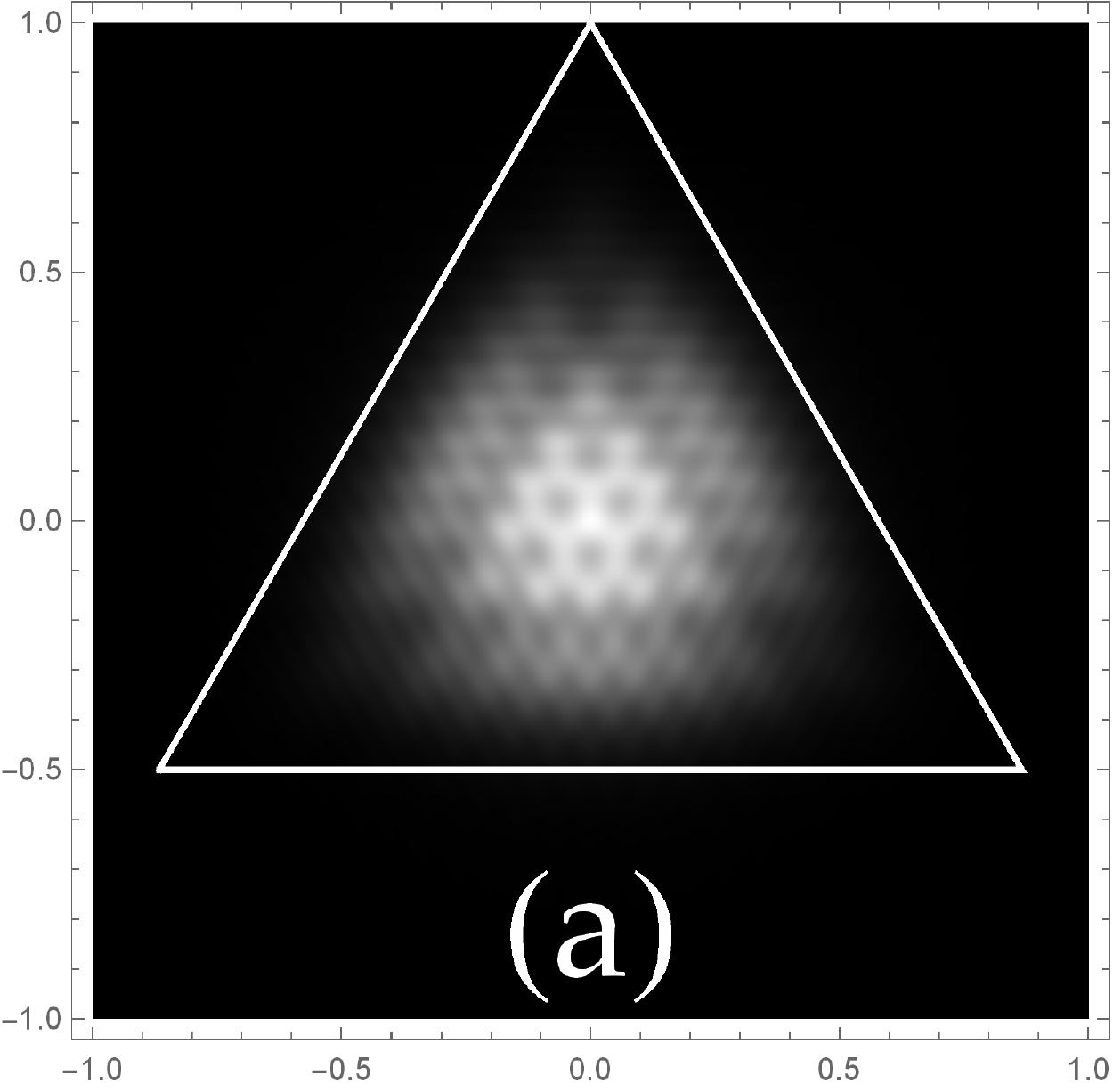}
\hfill
\includegraphics[width=4.8cm,angle=-0]{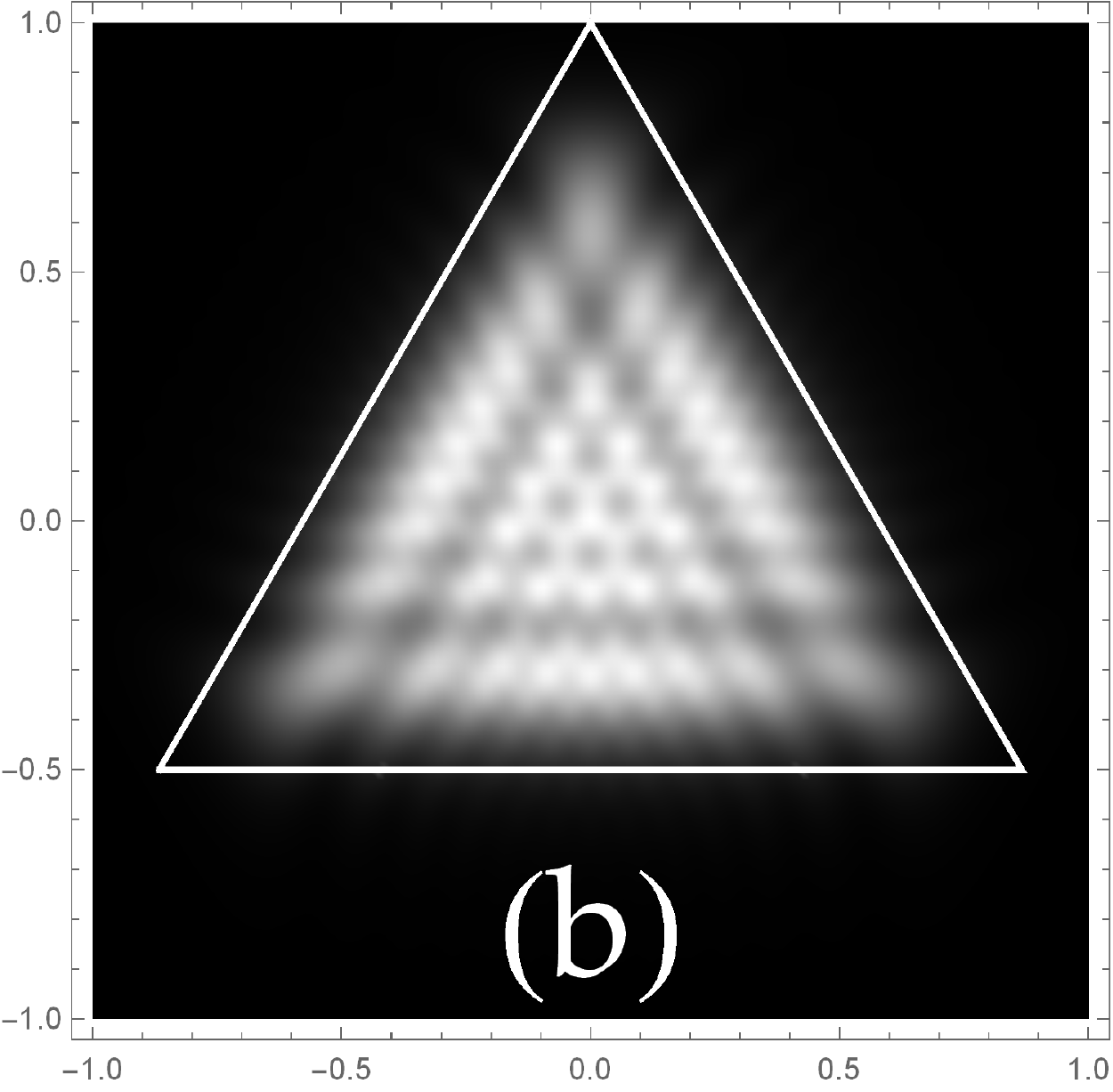}
\end{minipage}
\begin{minipage}[t]{10cm}
\includegraphics[width=4.8cm,angle=-0]{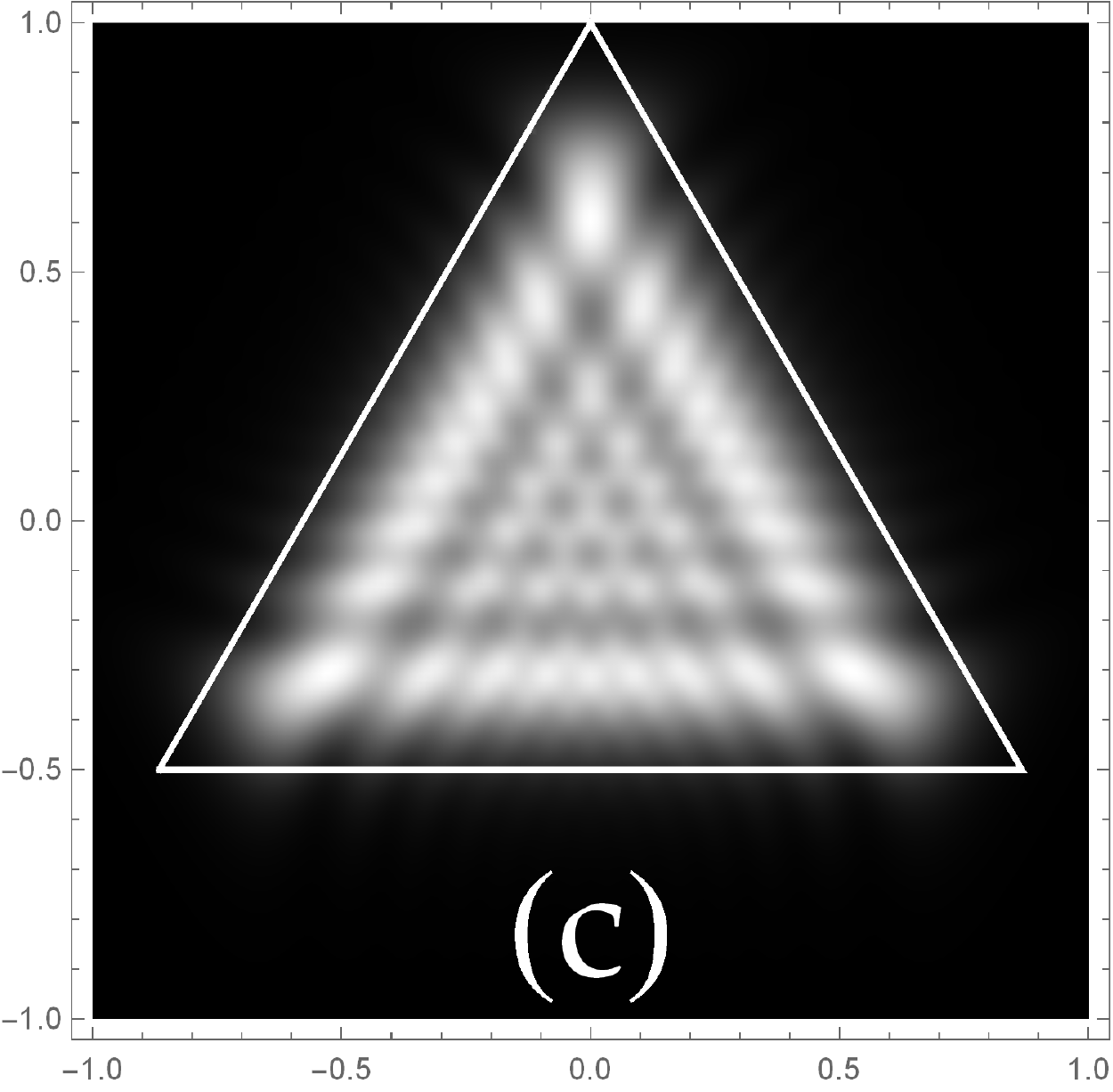}
\hfill
\includegraphics[width=4.8cm,angle=-0]{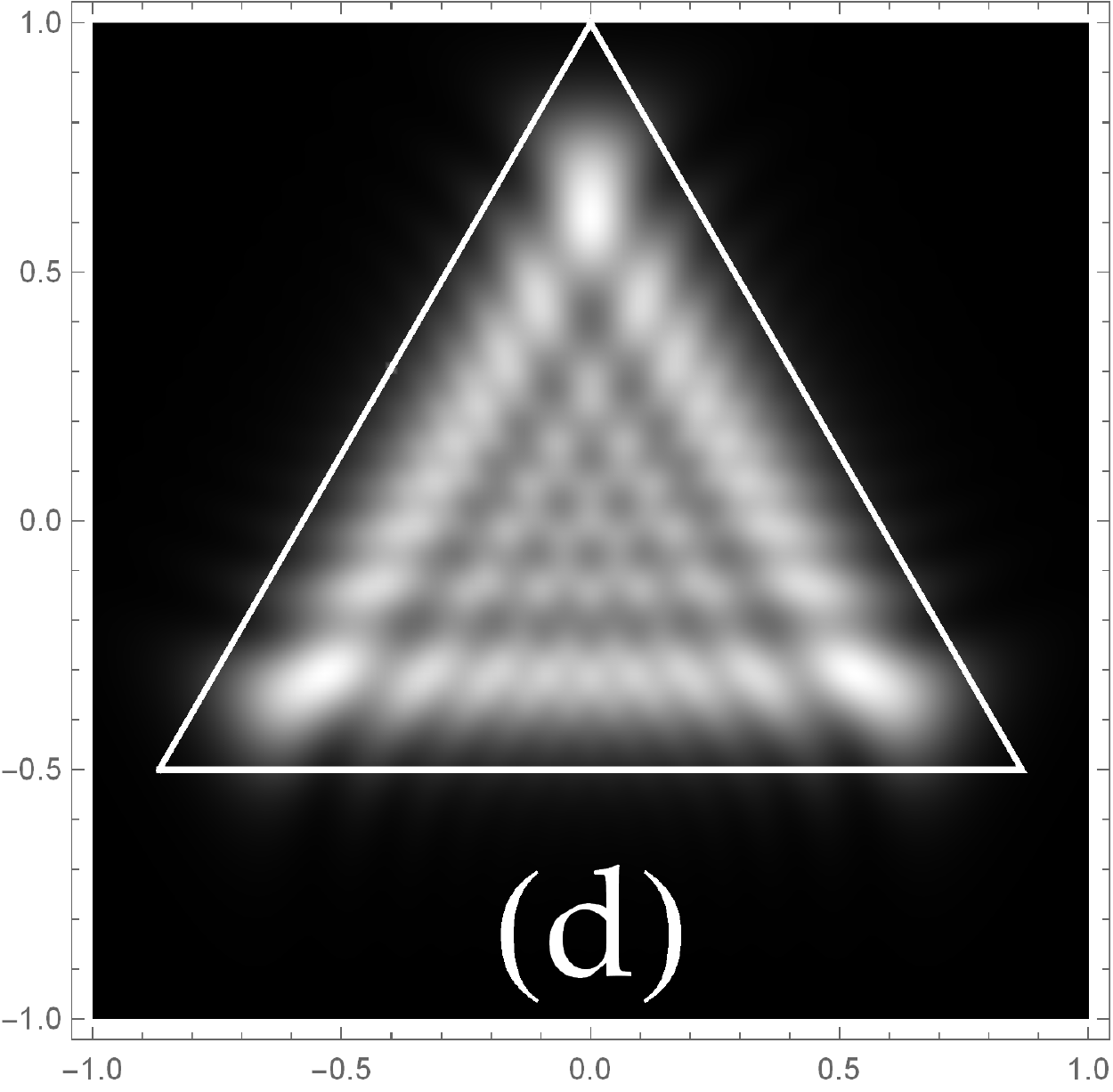}
\end{minipage}
\caption{Two-dimensional maps of the optical intensity produced by the diffraction of a Gaussian beam with $\lambda=0.5$~$\mu$m and $D=0$ at the 
observation plane $z=100$~mm. The radius of the circumscribed circle $\gamma_M$ is set to 1~mm. The spot-size has been set to $w_0=500$~$\mu$m (a), 
$w_0=1$~mm (b), $w_0=200$~$\mu$m (c), and $w_0=10$~mm (d).}
\label{Fig:BorghiConjecture.3}
\end{figure}

In order to offer at least one example of computation of $\psi_{\rm BDW}$, in Fig.~\ref{Fig:BorghiConjecture.3} two-dimensional maps of the optical intensity
produced by the diffraction of a Gaussian beam with $\lambda=0.5$~$\mu$m whose waist and aperture planes coincide ($D=0$), are shown at the 
observation plane $z=100$~mm. The radius of the circumscribed circle $\gamma_M$ is set to 1~mm, in order to reproduce Fig.~5a of~\cite{Huang/Christian/McDonald/2006}.
To put into evidence the influence of the Gaussian nature of the illumination, different values of the spot-size $w_0$ have been chosen, and precisely $w_0=500$~$\mu$m (a), 
$w_0=1$~mm (b), $w_0=200$~$\mu$m (c), and $w_0=10$~mm (d). 
In particular, Fig.~\ref{Fig:BorghiConjecture.3}d can be compared to Fig.~5a of~\cite{Huang/Christian/McDonald/2006}, which was generated by using, in place 
of our collimated Gaussian beam, a plane wave. Since in the present case the spot size is one order of magnitude greater than the triangle length, it should be expected the
Gaussian beam to act like a plane wave, as witnessed by the excellent agreement between the two figures.

\section{Summary and conclusions}
\label{Sec:Summary}

An  analytical exact derivation of the paraxial wavefield generated by arbitrarily shaped polygonal sharp-edge apertures under Gaussian
beam illumination has been presented. No further approximations beyond Fresnel diffraction have been invoked.
We believe our achievement could help in further improving the theoretical understanding of paraxial diffraction. 
Moreover, the analytical and complete knowledge of the diffracted field under such a realistic model of illumination should be acknowledged as an effective computational tool of 
considerable usefulness for a wide range of applications.


\newpage

\begin{thebibliography}{99}

\bibitem{Smith/Marsh/1974}
R. C. Smith and J. S. Marsh, 
``Diffraction patterns of simple apertures,'' 
J. Opt. Soc. Am. \textbf{64,} 798 - 803 (1974).

\bibitem{Komrska/1982}
J. Komrska, 
``Simple derivation of formulas for Fraunhofer diffraction at polygonal apertures,'' 
J. Opt. Soc. Am. \textbf{72, } 1382 - 1384 (1982).

\bibitem{Ganci/1984}
S. Ganci, 
``Simple derivation of formulas for Fraunhofer diffraction at polygonal apertures from Maggi - Rubinowicz
transformation,''
J. Opt. Soc. Am. A \textbf{1,} 559 - 561 (1984).

\bibitem{Forbes/Asatryan/1998}
G. W. Forbes and A. A. Asatryan,
``Reducing canonical diffraction problems to
singularity-free one-dimensional integrals,''
J. Opt. Soc. Am. A \textbf{15,} 1320 (1998).

\bibitem{Huang/Christian/McDonald/2006}
J. G. Huang, J. M. Christian, and G. S. McDonald,
``Fresnel diffraction and fractal patterns from polygonal apertures,''
J. Opt. Soc. Am. A
\textbf{23, }  2768 - 2774 (2006).

\bibitem{Naraga/Hermosa/2018}
J. Naraga and N. Hermosa,
``Diffraction of polygonal slits using catastrophe optics,''
J. Appl. Phys. \textbf{124,} 034902 (2018).

\bibitem{Born/Wolf/1999} M. Born and E. Wolf,
\emph{Principles of Optics}
(Cambridge University Press, Cambridge, 1999).

\bibitem{Miyamoto/Wolf/1962} K. Miyamoto and E. Wolf, 
``Generalization of the Maggi-Rubinowicz theory of the boundary-diffraction wave,'' 
J. Opt. Soc. Am. \textbf{52,} 615 
 -625 (part I) and 626-637 (part II) 
(1962).
 
\bibitem{Borghi/2015} R. Borghi,
``Uniform asymptotic of paraxial boundary-diffraction waves,''
J. Opt. Soc. Am. A \textbf{32,} 685 - 696 (2015).

\bibitem{Borghi/2016} R. Borghi,
``Catastrophe optics of sharp-edge diffraction,''
 Opt. Lett.  \textbf{41,} 3114 - 3117 (2016).

\bibitem{Borghi/2017} R. Borghi,
``Heart diffraction,''
 Opt. Lett.  \textbf{42,} 2070 - 2073 (2017).

\bibitem{Borghi/2018} R. Borghi,
``Tailoring axial intensity of laser beams with a heart-shaped hole,” by Wang et al.: Comment,''
 Opt. Lett.  \textbf{43,} 3240 (2018).

\bibitem{Borghi/2019} R. Borghi,
``Sharp-edge diffraction under Gaussian illumination: a paraxial revisitation of Miyamoto-Wolf’s theory,''
J. Opt. Soc. Am. A, \textbf{36,} 1048 (2019).

\bibitem{Owen/1956} D. B. Owen, 
Ann. Math. Statist. \textbf{27,}  1075 (1956).

\bibitem{Brychov/Savischenko/2016}
Yu. A. Brychkov and  N. V. Savischenko,
``Some properties of the Owen T-function,''
Integral Transforms and Special Functions  \textbf{27,}  163 
(2016)

\bibitem{Stahl/Gbur/2016}
C. Stahl and G. Gbur,
``Analytic calculation of vortex diffraction by a triangular aperture,''
J. Opt. Soc. Am. A \textbf{33,}1175 (2016).

\bibitem{Rivera/Galvin/Steinforth/Eden/2018}
J. A. Rivera, T.  C. Galvin, A.  W. Steinforth, and J. G. Eden
``Fractal modes and multi-beam generation from hybrid microlaser resonators,''
Nature Commun.  \textbf{9, }  2594 (2018).

\bibitem{Sroor/Naidoo/Miller/Nelson/Courtial/Forbes/2019}
H. Sroor, D.  Naidoo, S. W. Miller, J. Nelson, J. Courtial, and A. Forbes,
``Fractal light from lasers,''
Phys. Rev. A \textbf{99,}  013848  (2019).

\bibitem{Rocha/Amaral/Fonseca/Jesus-Silva/2019}
J. C. A. Rocha, J. P. Amaral, E. J. S. Fonseca, and A. J. Jesus-Silva
``Study of the conservation of the topological charge strength in diffraction by apertures ,''
J. Opt.  Soc. Am. B  \textbf{36,} 2114 (2019).

\bibitem{Otis/1974} G. Otis, 
``Application of the boundary-diffraction-wave theory to Gaussian beams,'' 
J. Opt. Soc. Am. \textbf{64,} 1545 - 1550 (1974).

\end{thebibliography}
\end{document}